\documentstyle[preprint,aps,epsf]{revtex}
\draft

\begin{document}

\title{Quantum Manipulations of Small Josephson Junctions}

\author {Alexander Shnirman$^{1,2}$, Gerd Sch\"on$^1$ and Ziv Hermon$^1$}

\address{
$^1$Institut f\"ur Theoretische Festk\"orperphysik,
Universit\"at Karlsruhe, D-76128 Karlsruhe, Germany. \\
$^2$ School of Physics and Astronomy, Tel Aviv University,
69978 Tel Aviv, Israel.}

\maketitle
\begin{abstract}
Low-capacitance Josephson junction arrays in the parameter range where
single charges can be controlled are suggested as possible physical
realizations of the elements which have been considered in the context 
of quantum computers. We discuss single and multiple quantum bit systems. 
The systems are controlled by applied gate voltages, which also allow the
necessary manipulation of the quantum states. We estimate that the
phase coherence time is sufficiently long for experimental demonstration
of the principles of quantum computation. 
\end{abstract}

\section{Introduction}
The issue of quantum computation has attracted much attention
recently \cite{Barenco_Review}. 
Quantum algorithms can perform certain types of calculations
much faster than classical computers~\cite{Shor_Factorization}. 
The basic concepts of quantum computation are 
quantum operations (gates) on quantum bits (qubits) and  registers, 
i.\ e.\ arrays of qubits. A qubit can be any two-level system 
which can be prepared in arbitrary superpositions of
its two eigenstates. Quantum computation requires
``quantum state engineering'', i.\ e.\ the processes of
preparation and manipulation
of these quantum states in a controllable way.
For quantum registers also ``entangled states'' (like the 
EPR state of two spins) have to be constructed. This necessitates
a coupling between different qubits, and one has to be able to
construct many-qubits  states (many-particle states) in a controllable way. 
Several physical systems have been proposed as qubits 
\cite{Ion_Chain_Zoller,Loss_Grains}. 
A serious limitation is the requirement that the phase coherence time
is sufficiently long to allow the coherent quantum manipulations.
The theoretically and experimentally  most advanced 
system appears to be the chain with trapped ions~\cite{Ion_Chain_Zoller}.

In this paper we propose an alternative system, composed
of low-capacitance Josephson junctions. The coherent 
tunneling of Cooper pairs mixes different charge states. 
By controlling the gate voltages we can control the strength of the
mixing. The physics of coherent Cooper-pair tunneling in this system
has been established before~\cite{Maassen,Tuo,Siewert}. The algorithms
of quantum computation introduce new, well-defined rules. Their
realization in experiments creates a new challenge.
Here, we consider a one-bit system
and we describe the possible ways of constructing quantum states.
Then we focus on a two-bit system, where we propose a controllable
coupling and we discuss the  construction of two-bit states.
In Section 2, we consider ideal systems. In Section 3 we include the
coupling to a realistic  external electrodynamic environment
which limits the phase coherence time.

\section{Idealized one and two-bit systems}
\subsection{One-bit system and gates.}   
\begin{figure}
\epsfysize=10\baselineskip
\centerline{\hbox{\epsffile{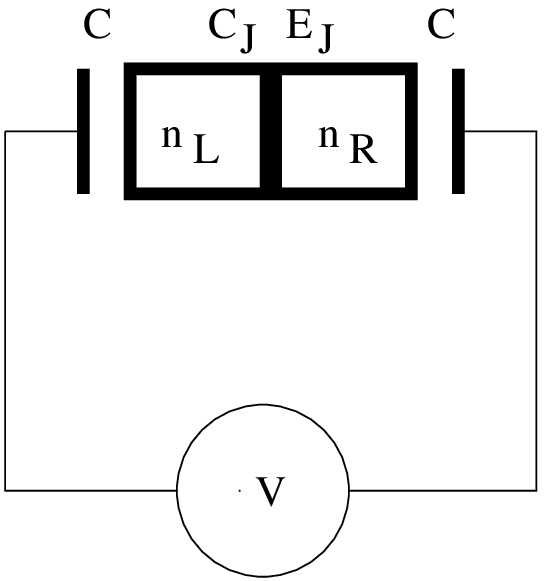}}}
\vskip 0.8cm
\caption[]{\label{ONE_BIT_IDEAL}
An idealized qubit system.}
\end{figure}

The system we propose as a qubit is shown in Fig.\,\ref{ONE_BIT_IDEAL}.
It consists of two small superconducting grains connected by a tunnel
junction. The junction is characterized by its capacitance $C_{\rm J}$ and
the Josephson energy $E_{\rm J}$. An ideal voltage source is connected
to the system via two external capacitors $C$. 
We assume the total number of electrons in both grains to be 
even, so that at low enough temperatures all  electrons are paired.
Moreover, we assume that the superconducting energy gap $\Delta$ 
is the largest energy in the problem. Thus we can neglect 
pair-breaking processes. The complete set of possible quantum states is,
then, characterized by $n_L$ and $n_R$ - the numbers of extra Cooper
pairs on the left and the right island, respectively. Since there
is no tunneling through the external capacitors, the total number
of the extra Cooper pairs $N \equiv n_L + n_R$ is constant.
In the following we assume $N=0$.
Hence the set of basis states is parameterized by the number of 
Cooper pairs on one island or their
difference $n \equiv (n_L - n_R)/2$. 
The Hamiltonian of this system is
\begin{equation}
\label{1bit_Hamiltonian}
        H = {(2en - {C V/2})^2\over{C+2C_{\rm J}}} - 
        E_{\rm J} \cos(2\pi {\Theta \over {\Phi_0/2}}) \ ,
\end{equation}
where $\Theta$ is the conjugate to the variable $2en$ 
\cite{Carruthers_Phase_Operator} and  
$\Phi_0 = {h\over e}$ is the flux quantum.
We consider systems where the charging energy of the internal
capacitor $E_{C_{\rm J}} = (2e)^2/2C_{\rm J}$ is much larger than 
$E_{\rm J}$.
In this regime the energy levels are determined by the
charging part of the Hamiltonian (\ref{1bit_Hamiltonian})
for most values of the external voltage $V$. However,
when $V$ is such that the charging energies of two 
neighboring states $|n>$ and $|n+1>$ are close to each other, the Josephson
tunneling becomes relevant. Then the two  eigenstates are symmetric
and antisymmetric combinations of $|n>$ and $|n+1>$ with an energy
gap of $E_{\rm J}$ between them. 

We concentrate on the voltage interval near one such degeneracy point.
We demand, however, the interval to be wide enough, so that for most
values of the voltage away from degeneracy to a reasonable approximation 
the eigenstates are $|n>$ 
and $|n+1>$. Near the degeneracy points
 we can translate
(\ref{1bit_Hamiltonian}) to the spin-$1\over2$ language:
\begin{equation}
\label{Magnetic_Hamiltonian}
H= {eCV\over {C+2C_{\rm J}}} \sigma_z + {E_{\rm J}\over 2} \sigma_x \ , 
\end{equation}
where $|\uparrow>\equiv |n>$ and $|\downarrow>\equiv |n+1>$.     
Using this language we can immediately propose a few 
one-bit operations. If, for example, one chooses 
the operating point (i.\ e.\ the voltage) away from the degeneracy, 
the logical states $|0>$ and $|1>$
are just $|\downarrow>$ and $|\uparrow>$ respectively. Then, switching 
the system suddenly to the degeneracy point for a time
$\Delta t$ and switching suddenly back, we can perform one of the basic 
one-bit operations (gates) - a spin flip:
\begin{equation}
\label{1bit_spin_flip}
U_{\rm flip}(\Delta t)=  
\left( 
\begin{array}{cc} \cos({E_{\rm J} \Delta t\over 2\hbar})  & 
                       i\sin({E_{\rm J} \Delta t\over 2\hbar}) \\ 
                  i\sin({E_{\rm J} \Delta t\over 2\hbar}) & 
                       \cos({E_{\rm J} \Delta t\over 2\hbar}) 
\end{array} 
\right)   
\end{equation}
To get rid of the time-dependent phases we perform all the calculations
is the interaction picture, when the zero order Hamiltonian is 
the Hamiltonian
at the operation point. To estimate the time-width $\Delta t$
of the voltage pulse
needed for a total spin flip, we note that a reasonable  experimental
value of $E_{\rm J}$ is of the order of $1$K. It cannot 
be chosen much smaller,
since the condition $k_{\rm B} T \ll E_{\rm J}$ must be satisfied. Therefore 
the time-width is very short: $\Delta t \approx 10^{-10}$s. 

However there exists an alternative way to perform a coherent spin flip, 
as follows:
The system is pushed adiabatically to the degeneracy point. Then the
ac voltage with frequency $E_{\rm J}/\hbar$ is applied. The process is
completely analogous to the paramagnetic resonance (here the constant
magnetic field component is in the x-direction, while the oscillating 
one is in the z-direction). The time-width of the ac pulse needed 
for the total spin flip depends on its amplitude, therefore it can be 
chosen much longer then $10^{-10}$s.

One can also perform an operation which changes only the relative
phase between the the states $|\downarrow>$ and $|\uparrow>$.
To this end one should suddenly switch to another voltage away
from the degeneracy and switch suddenly to the original point. 
The phase is acquired due to difference in energy gaps between the 
two points.       

\subsection{Two-bit system.}
\begin{figure}
\epsfysize=8\baselineskip
\centerline{\hbox{\epsffile{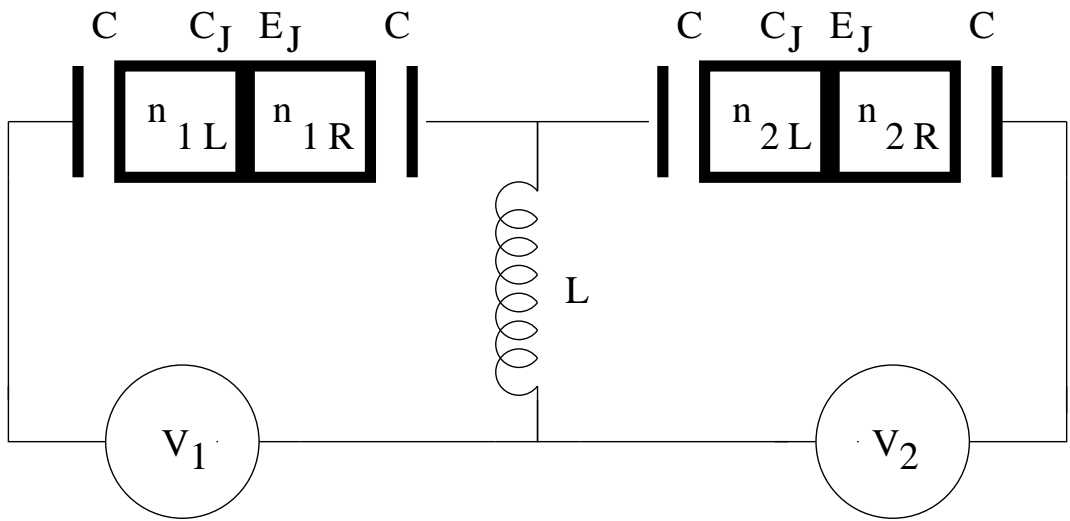}}}
\vskip 0.8cm
\caption[]{\label{TWO_BIT_IDEAL}
An idealized two-bit system.}
\end{figure}

To perform two-bit operations (gates) which result in entangled
states, one has to couple the qubits in a controllable way.
The best situation would be one where the coupling can be 
switched off, leaving the qubits uncoupled when no operations
are performed. Unfortunately, this option appears difficult 
to realize
in microscopic and mesoscopic systems. As an alternative we suggest
for a system with a weak constant coupling between the qubits. We will
show that by tuning the energy gaps of the individual qubits 
we can change the effective strength of the coupling. We propose to
couple two qubits using an inductance as shown in Fig.(\ref{TWO_BIT_IDEAL}).
One can easily see that if $L=0$ the system reduces to two
uncoupled qubits, while if $L=\infty$ the Coulomb interaction
couples both qubits strongly. Therefore, we are interested
in low values of $L$ (the exact criterion to be specified).
The Hamiltonian describing this system is
\begin{eqnarray}
\label{2bit_Hamiltonian}
H = &&  \sum_{i=1,2} \left\{
{(2e n_i - V_i C_{\rm t})^2\over 2C_{\rm J}} - 
E_{\rm J} \cos(2\pi {\Theta_i \over {\Phi_0/2}})
\right\} + 
{q^2\over 2(2C_{\rm t})} + {\phi^2\over 2L} \nonumber \\ - 
&& {e(n_1-n_2)q\over C_{\rm J}} - 
{e^2 C_{\rm t} \over C_{\rm J}^2}(n_1+n_2)^2  \ ,
\end{eqnarray}
where $q$ denotes the total charge on the external capacitors of 
both qubits, $\phi$ is its conjugate variable, 
and  $C_{\rm t}^{-1}= C_{\rm J}^{-1} + 2 C^{-1}$.
The oscillator described by $q$ and $\phi$ variables produces 
an effective mean field coupling between
the qubits for frequencies smaller than 
$\omega_{LC} = {1/\sqrt{2C_{\rm t} L}}$. In order to have this coupling in 
a wide enough voltage interval around the degeneracy point, we demand that
\begin{equation}
\label{High_Frequency_Condition}
A\equiv{\hbar\omega_{LC}\over E_{\rm J}}\gg 1\ .
\end{equation}
Although the integration over $q$ and $\phi$ can be performed 
exactly starting with (\ref{2bit_Hamiltonian}), the resulting 
effective action is not simple. We prefer an alternative 
way. First we perform a canonical transformation
\begin{equation}
\label{Canonical_Transformation_2bit}
\begin{array}{ll}
   \tilde q = q - {2e(n_1-n_2)C_{\rm t}\over C_{\rm J}} \ ,&  
              \ \ \ \tilde \phi = \phi \ ;\\
   \tilde n_i = n_i  \ , &  \ \ \ \tilde \Theta_i = \Theta_i - 
   (-1)^i \, {C_{\rm t}\over C_{\rm J}}\phi \ ,
   \end{array} 
\end{equation}   
$i=1,\,2$, which results in a new Hamiltonian (we omit the tildes):
\begin{eqnarray}
\label{2bit_Hamiltonian_Cosine}
H = && \sum_{i=1,2} \left\{
{(2e n_i - {CV_i/2})^2\over{C+2C_{\rm J}}} - 
E_{\rm J} \cos\left[{2\pi\over{\Phi_0/2}} \left(\Theta_i + (-1)^i \,
{C_{\rm t}\over C_{\rm J}} \phi\right)\right] 
\right\} \nonumber \\ 
&& + {q^2\over 2(2C_{\rm t})} + {\phi^2\over 2L} \ .
\end{eqnarray}
Now we make an additional assumption
\begin{equation}
\label{Small_Fluctuation_Assumption}
<\phi^2> \, \ll \,  
\left({C_{\rm J}\over C_{\rm t}}{\Phi_0\over2}\right)^2 \ . 
\end{equation}
This assumption is, actually, necessary if one wishes to
have Josephson tunneling terms in the Hamiltonian.
Qualitatively, one can observe from (\ref{2bit_Hamiltonian_Cosine}),
that if (\ref{Small_Fluctuation_Assumption}) does not hold,
the $E_{\rm J} \cos(..)$ terms are washed out. Below we will obtain this
result in a more rigorous way. 

Assuming the condition (\ref{Small_Fluctuation_Assumption})
we expand the $E_{\rm J} \cos(..)$ terms of (\ref{2bit_Hamiltonian_Cosine}) 
in powers of $\phi$ and we neglect powers higher than linear.
Then we can perform the integrations over $q$ and $\phi$, which
are again Gaussian. As a result we obtain an effective Hamiltonian,
which consists of two one-bit Hamiltonians (\ref{1bit_Hamiltonian})
and a coupling term:
\begin{equation}
\label{2bit_Coupling}
H_{\rm int} = E_L
\left[
\sin\left({2\pi\over{\Phi_0/2}} \Theta_1\right) - 
\sin\left({2\pi\over{\Phi_0/2}} \Theta_2\right)
\right]^2 \ ,
\end{equation}     
where
\begin{equation}
\label{E_L}
E_L = 8\pi^2 {C_{\rm t}^2\over C_{\rm J}^2} {E_{\rm J}^2 L\over\Phi_0^2}\ .
\end{equation}
This Hamiltonian is valid for frequencies which are smaller than
$\omega_{LC}$.
In the spin-$1\over2$ language it is
\begin{equation}
\label{2bit_Spin_Coupling}
H_{\rm int} = -{E_L\over 4} (\sigma_y^{(1)} - \sigma_y^{(2)})^2 \ . 
\end{equation}
The mixed term in (\ref{2bit_Spin_Coupling}) is
important in certain situations. Assume that 
$E_L$ is comparable to $E_{\rm J}$ and the voltages $V_1$ and
$V_2$ are such that both qubits are out of degeneracy.
Then, to a good approximation, the eigenstates of the 
two-bit system without coupling are $|\downarrow\downarrow>$,
$|\downarrow\uparrow>$, $|\uparrow\downarrow>$ and
$|\uparrow\uparrow>$. In a general situation,
these states are separated by energies which are
larger or much larger than $E_{\rm J}$. Therefore, 
the effect of the coupling is small. If, however, a pair of these
state is degenerate, the coupling may lift the degeneracy,
changing the eigenstates drastically. For example, if $V_1 = V_2$,
the states $|\downarrow\uparrow>$ and $|\uparrow\downarrow>$
are degenerate. In this case the correct eigenstates are:
${1\over\sqrt{2}}(|\downarrow\uparrow> + |\uparrow\downarrow>)$  
and
${1\over\sqrt{2}}(|\downarrow\uparrow> - |\uparrow\downarrow>)$
with the energy splitting $E_L$ between them. 

Now we are ready to propose a way to perform two-bit operations
which will result in entangled states. For this we choose the
operating points for the qubits at different voltages. Then we
switch suddenly the voltages to be equal for a time $\Delta t$ and  
switch suddenly back. The result is a ``generalized'' 
spin-flip, which may be described in the basis 
\{$|\downarrow\downarrow>$,
$|\downarrow\uparrow>$, $|\uparrow\downarrow>$,
$|\uparrow\uparrow>$\}
by a matrix:
\begin{equation}
\label{2bit_Operation}
U_{\rm flip}^{(2)}(\Delta t)=  
\left( 
\begin{array}{cccc}
1 & 0 & 0 & 0 \\ 
0 & \cos\left({E_L \Delta t\over 2\hbar}\right)  & 
i\sin\left({E_L \Delta t \over 2\hbar}\right) & 0 \\ 
0 & i\sin\left({E_L \Delta t \over 2\hbar}\right) & 
\cos\left({E_L \Delta t\over 2\hbar}\right)  & 0 \\
0 & 0 & 0 & 1 
\end{array} 
\right) \ .
\end{equation}
The gate (\ref{2bit_Operation}) together with the
one-bit gates (\ref{1bit_spin_flip}) constitute a universal set,
i.e., they are sufficient for quantum computation.

Instead of applying very short voltage pulses,
one again can move the system adiabatically to the
degeneracy point ($V_1=V_2$), and, then, apply an ac
voltage pulse in the antisymmetric channel 
$V_1 - V_2 = A \exp(i E_L t/\hbar)$.

\section{Circuit effects and dissipation}
\subsection{One-bit system. Energy relaxation out of the degeneracy point.}
\begin{figure}
\epsfysize=10\baselineskip
\centerline{\hbox{\epsffile{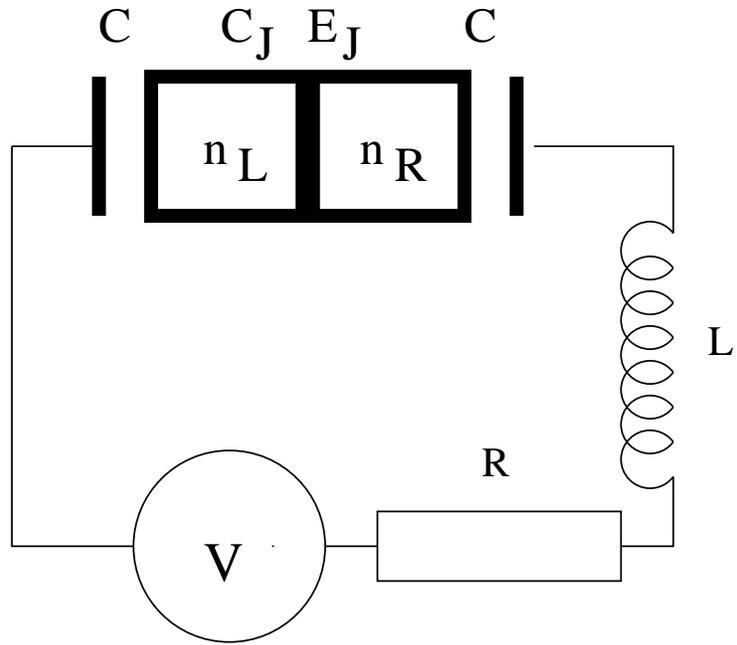}}}
\vskip 0.8cm
\caption[]{\label{ONE_BIT_DISSIPATION}
A qubit with dissipation.}
\end{figure}

The idealized picture outlined above has to
be corrected with respect to the possible dissipation mechanisms,
which, as usual, cause decoherence and energy relaxation.
In this paper we focus on the effect of ohmic dissipation in
the circuit, which originates mostly from the voltage sources.
We also consider the effect of LC resonances in the circuit. 
We neglect quasi-particle tunneling since it is strongly suppressed at
low temperatures and low voltages.
The system is shown in Fig.\,\ref{ONE_BIT_DISSIPATION}.
We include the inductance $L$ explicitly for two reasons.
First, we would like to couple the bath of oscillators to a continuous
charge variable, rather than directly coupling them to the discrete 
variable $n$. Second, we would like to treat the $LC$ oscillatory mode
separately from the bath oscillators, since it plays an important
role in the two-bit system. The Hamiltonian for this system is:
\begin{equation}
\label{1bit_Dissipative_Hamiltonian}
H = {(2en - VC_{\rm t})^2\over 2C_{\rm J}} - 
E_{\rm J} \cos\left(2\pi {\Theta \over {\Phi_0/2}}\right) +
{q^2\over 2C_{\rm t}} + {\phi^2\over 2L} - {2enq\over C_{\rm J}} + 
H_{\rm bath}(q) \ ,
\end{equation} 
where
\begin{equation}
\label{Bath_Hamiltonian_q}
H_{\rm bath}(q) = \sum_j \left[ 
{p_j^2\over {2m_j}} + 
{m_j \omega_j^2\over 2}\left(x_j - 
{\lambda_j\over{m_j \omega_j^2}} q\right)^2 \right] \ ,
\end{equation} 
and
$J(w) \equiv {\pi\over 2} \sum_j {\lambda_j^2\over {m_j 
\omega_j}} \delta(\omega - \omega_j) = R\,\omega $ .
In (\ref{1bit_Dissipative_Hamiltonian}) and (\ref{Bath_Hamiltonian_q}) 
the variable $q$ denotes the charge on the external capacitors, 
while $\phi$ is its conjugate variable. The coupling to the bath
(\ref{Bath_Hamiltonian_q}) produces the usual
dissipative term - $R\dot q$ - in the equation of motion for $q$.

Although the Hamiltonian (\ref{1bit_Dissipative_Hamiltonian}) is quite
transparent, another equivalent form is more convenient. 
It is obtained after two consequent canonical transformations:
\begin{equation}
\label{Canonical_Transformation_phi}
\begin{array}{ll}
   \tilde q = q\ ,  &  \ \ \ \tilde \phi = \phi + 
   \sum_{j} {\lambda_j\over{m_j\omega_j^2}} p_j\ ;  \\
   \tilde x_j = x_j-
   {\lambda_j\over{m_j\omega_j^2}} q\ ,  &  
   \ \ \ \tilde p_j = p_j\ ,
  \end{array}  
\end{equation}
and
\begin{equation}
\label{Canonical_Transformation_Theta}
\begin{array}{ll}
\bar q = \tilde q - {2enC_{\rm t}\over C_{\rm J}}\ ,  
& \ \ \ \bar \phi = \tilde \phi\ ; \\
\bar n = n\ ,  &  \ \ \ \bar \Theta = 
\Theta + {C_{\rm t}\over C_{\rm J}}\tilde \phi\ .
\end{array}
\end{equation}
Now the Hamiltonian takes the form:
\begin{equation}
\label{1bit_Dissipative_Hamiltonian_phi}
H = {(2en - {CV/2})^2\over{C+2C_{\rm J}}} - 
E_{\rm J} \cos\left[{2\pi\over{\Phi_0/2}} \left(\bar\Theta - 
{C_{\rm t}\over C_{\rm J}}\tilde\phi\right)\right] +
{\bar q^2\over 2C_{\rm t}} + H'_{\rm bath}(\tilde\phi) \ .
\end{equation}
where
\begin{equation}
\label{Bath_Hamiltonian_phi}
H'_{\rm bath}(\tilde \phi) = \sum_j \left[ 
{\tilde p_j^2\over {2m_j}} + 
{m_j \omega_j^2\over 2} \tilde x_j^2 \right] +  
{(\tilde \phi - \sum_j {\lambda_j\over{m_j 
\omega_j^2}} \tilde p_j)^2 
\over {2L}}  \ .
\end{equation}
The two forms  (\ref{1bit_Dissipative_Hamiltonian}) and
(\ref{1bit_Dissipative_Hamiltonian_phi}) differ mostly
in the definition of the phase variable conjugated to $q$.
If $L$ is small enough, the variable $\phi$ fluctuates
in a small interval around zero. However, the fluctuations of $\tilde\phi$ 
are around a bath-dependent value, which may fluctuate strongly itself. 
Below we will see that both forms may be useful in different
situations.   

First, we estimate the energy relaxation time, $\tau_r$, due to the
ohmic dissipation. We assume that the system is away from
the degeneracy point and is prepared in one of its eigenstates
($|n>$ or $|n+1>$) at the beginning. We use the Hamiltonian in the 
second form (\ref{1bit_Dissipative_Hamiltonian_phi}). 
The part of (\ref{1bit_Dissipative_Hamiltonian_phi}) 
connecting the states $|n>$ and $|n+1>$
is 
\begin{equation}
\label{Tunneling_Hamiltonian}
H_{\rm t} = {E_{\rm J}\over2} 
\exp\left(i{2\pi\over{\Phi_0/2}}\bar\Theta\right) 
\exp\left(-i{2\pi\over{\Phi_0/2}} 
{C_{\rm t}\over C_{\rm J}}\tilde\phi\right) + {\rm h.c.} \ .
\end{equation}
At this point we can directly
apply the standard Golden Rule results for the transition
rate from the state $|n>$ to the state $|n+1>$
\cite{P(E)_Panyukov_Zaikin,P(E)_Odintsov,P(E)_Nazarov,P(E)_Devoret} :
\begin{eqnarray}
\label{Gamma}
\Gamma(\Delta E) = && {\pi\over {2\hbar}} E_{\rm J}^2 P(\Delta E) \ , \\ 
\label{P(E)}
P(\Delta E) = && {1\over{2\pi\hbar}}
\int_{-\infty}^{\infty} dt \  \exp\left[4{C_{\rm t}^2\over C_{\rm J}^2} K(t) + 
{i\over\hbar}\Delta E t\right] \ , \\ 
K(t) = &&  2\int_0^{\infty} {d\omega \over\omega} 
{Re Z_{\rm t}(\omega)\over R_K}
\left[\coth\left({\hbar\omega\over2 k_B T}\right)[\cos(\omega t)-1]
-i\sin(\omega t)
\right] \ , \\ 
Z_{\rm t}^{-1} = && i\omega C_{\rm t} + {1\over{R + i\omega L}} \ .
\end{eqnarray}
Here $\Delta E$ is the energy gap between the two states.
The qualitative behavior of the system is, as usual, controlled
by the dimensionless conductance $g = {R_{\rm K}\over 4R}$. In our system
a remarkable renormalization of the controlling parameter $g$ occurs: from 
(\ref{P(E)}) one can observe that 
$\tilde g = {C_{\rm J}^2\over C_{\rm t}^2} g$ is the relevant parameter. 
Thus, making the external
capacitances $C$ much smaller than the internal one - $C_{\rm J}$, we
can drastically reduce the effect of the dissipation.
Physically, this means that the fluctuations produced
by the resistor are screened by the small capacitors, and, therefore,
they have little effect on the Josephson junction. 

To be more concrete, we exploit the asymptotic formula for 
$P(\Delta E)$ \cite{P(E)_Devoret}
\begin{equation}
\label{P(E)_Asymptotic}
P(\Delta E) = {{\exp(-2\gamma/\tilde g)}\over {\Gamma(2/\tilde g)}}
{1\over{\Delta E}} 
\left[ {\pi\over \tilde g} 
{\Delta E \over E_{C_{\rm t}}} \right]^{2/\tilde g} \ , 
\end{equation}
where $\Gamma(..)$ is the Gamma function.
For large values of $\tilde g$ we obtain:
\begin{equation}
\label{Energy_Relaxation_Time}
\tau_r \equiv {1\over \Gamma(\Delta E)} \approx
\tau_{\rm op} 
{\tilde g \over 2\pi^2} 
{\Delta E\over E_{\rm J}} \ ,
\end{equation}
where $\tau_{\rm op} \approx {h \over E_{\rm J}}$ is the operation time
(see (\ref{1bit_spin_flip})). 

\subsection{One-bit system. Circuit effects at the degeneracy point.}
At the degeneracy point the system is equivalent to the 
two-level model with a weak ohmic dissipation, which was 
extensively studied during the last decades \cite{Two_Level_Leggett}.
It is well known that when $\tilde g \gg 1$ coherent 
oscillations take place. These oscillations make the spin-flip operation
(\ref{1bit_spin_flip}) possible. 
The decay time of the coherent oscillations
is given by
\begin{equation}
\label{TAU_D} 
\tau_{d} \approx {\tilde g\over 2\pi^2} 
{h\over E_{\rm J}} = {\tilde g \over 2\pi^2} \tau_{op} \ ,
\end{equation} 
and the energy gap $E_{\rm J}$ is slightly renormalized:
\begin{equation}
\label{GAP_RENORMALIZATION} 
E_{\rm J} \rightarrow E_{\rm J} 
\left({E_{\rm J}\over \hbar\omega_c}\right)^{1\over{g-1}} \ .   
\end{equation}
The physical cut-off $\omega_c$ is usually a system-dependent property.
For a pure ohmic dissipation caused by a metallic resistor it
may be as high as the Drude frequency. However, when additional
capacitances and inductances are present in the circuit,
the relevant cut-off may be lowered to the characteristic
$LC$ frequencies. 

As was said above, the $LC$ phase fluctuations can wash out the 
Josephson coupling.
To see this, we 
begin with the first form of the Hamiltonian 
(\ref{1bit_Dissipative_Hamiltonian}) and we integrate
out the bath variables and the oscillatory mode variables - $\phi , q$.
The partition function reads:
\begin{eqnarray}
\label{Partition_Function}
&& Z= \sum_{n_0} \int_{n_0}^{n_0} DnD\Theta \times \nonumber \\ 
&& \exp\left\{ {1\over \hbar} 
 \left[\int_0^{\hbar \beta} d\tau (2ei\Theta\dot n - H_{0}(n,\Theta)) 
 - \int_0^{\hbar\beta} \int_0^{\hbar\beta} d\tau d\tau' 
{1\over2} G(\tau-\tau') n(\tau)n(\tau')\right] \right\}   \ ,
\end{eqnarray} 
where
\begin{equation}
\label{H_0}
H_0 = {(2en - VC_{\rm t})^2\over 2C_{\rm J}} - 
E_{\rm J} \cos\left(2\pi {\Theta \over {\Phi_0/2}}\right) \ ,
\end{equation}
and
\begin{equation}
\label{G(TAU_TAU)}
G(\omega_n) = -{4 C_{\rm t} e^2\over C_{\rm J}^2} \ 
{1\over{(1 + C_{\rm t} L \omega_n^2 + C_{\rm t} R |\omega_n|)}} \ .  
\end{equation}
Below we will show, that in the relevant parameters' range,
the following inequality holds
\begin{equation}
\label{Frequences_Unequality}
{1\over C_{\rm t} R} \gg {1\over\sqrt{L C_{\rm t}}} \gg {R\over L} \ .
\end{equation}
Therefore, the natural cutoff for (\ref{G(TAU_TAU)}) would be
$\omega_c = \omega_{LC} = {1\over\sqrt{L C_{\rm t}}}$. 
To make a rough estimation we replace (\ref{G(TAU_TAU)}) by
\begin{eqnarray}
\label{G_APPROXIMATE}
G(\omega_n) = && -{4 C_{\rm t} e^2\over C_{\rm J}^2}
(1 - C_{\rm t} L \omega_n^2 - C_{\rm t} R) \ , 
\ \ \omega_n < \omega_c \nonumber \\
G(\omega_n) = && 0 \ , \ \ \omega_n > \omega_c \ .
\end{eqnarray}
The kernel (\ref{G_APPROXIMATE}) is a sum of a trivial
charging energy renormalization (the first term), the usual
ohmic dissipation (the third term) and the inductive term,
which we will focus on.
We apply the standard charge representation technique
\cite{Schon_Zaikin_Review}, expanding 
$\exp\left[{1\over \hbar} \int_0^{\hbar \beta}
d\tau E_{\rm J} \cos(2\pi {\Theta \over {\Phi_0/2}}) \right]$ in powers of 
$E_{\rm J}$ 
and integrating over $\Theta$ term by term. One obtains, then,
a path integral over integer charge paths with instantaneous ``jumps'' 
between the different values of $n$. Each such ``jump'' contributes
a multiplicative factor of ${E_{\rm J}/2\hbar}$ to the weight of the path.
It turns out that the inductive term contributes another multiplicative
factor for each ``jump'', so that $E_{\rm J}$ is renormalized as:
\begin{equation}
\label{E_J_RENORMALIZATION}
\tilde E_{\rm J} = E_{\rm J}\, \exp\left({-{2LC_{\rm t}^2\omega_c 
\over{\pi R_K C_{\rm J}^2}}}\right) \ .
\end{equation} 
One can immediately observe that the condition that $E_{\rm J}$ is not
renormalized to zero coincides with the small fluctuations condition
(\ref{Small_Fluctuation_Assumption}). We would like to emphasize
that the phase fluctuations which may wash out the Josephson coupling
in (\ref{E_J_RENORMALIZATION})
are related to the ``weakly fluctuating'' phase $\phi$, rather than
to the ``strongly fluctuating'' $\tilde \phi$ 
(see (\ref{Canonical_Transformation_phi})). Thus the effects
of the inductance and the dissipation may be well separated in this regime.  
Another way of viewing this separation is by noting that the $LC$ phase
fluctuations are fast, therefore they effectively wash out the slower
processes (like Josephson tunneling). On the other hand the phase
fluctuations, caused by the resistor, are large only at low frequencies,
thus they cannot wash out the faster processes.

\subsection{Two-bit system. Circuit effects.}
\begin{figure}
\epsfysize=8\baselineskip
\centerline{\hbox{\epsffile{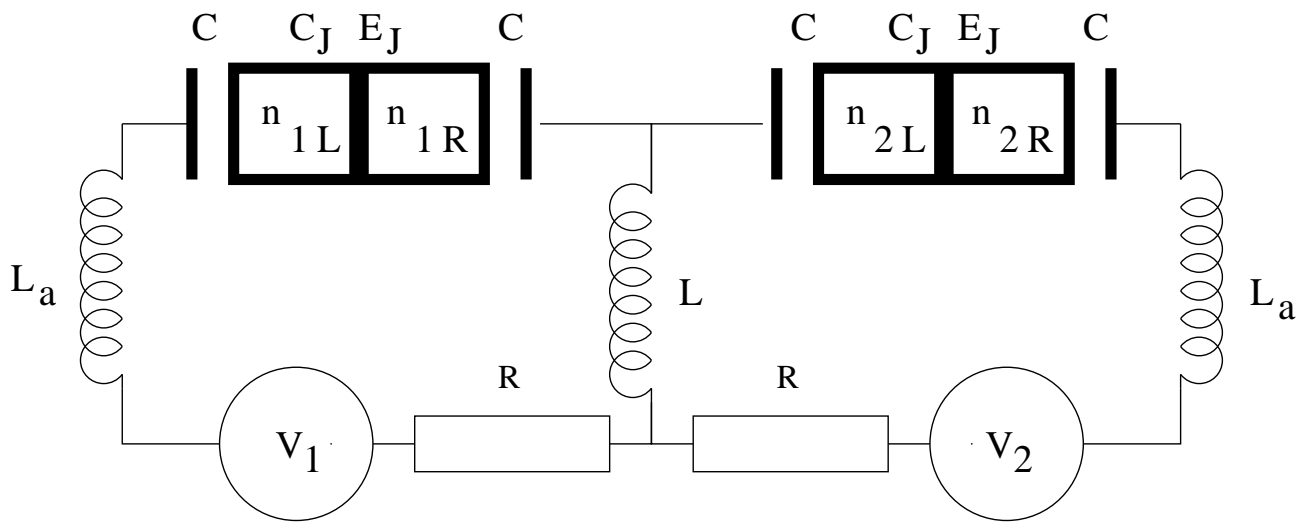}}}
\vskip 0.8cm
\caption[]{\label{TWO_BIT_DISSIPATION}
Two-bit system with dissipation.}
\end{figure}

In this subsection we show that the coupling term (\ref{2bit_Coupling}) 
survives
when the ohmic dissipation is present in the circuit. The analyzed
system is shown in Fig.(\ref{TWO_BIT_DISSIPATION}). We include
the auxiliary inductances $L_a$ in order to couple the 
dissipation bathes to continuous charge variable, rather than
directly to the principal variables $n$ and $m$. Finally
we take $L_a \rightarrow 0$. The Hamiltonian is given by:
\begin{eqnarray}
\label{2bit_Dissipative_Hamiltonian}
H = && \sum_{i=1,2} \left\{
{(2e n_i - V_i C_{\rm t})^2\over 2C_{\rm J}} - E_{\rm J} 
\cos\left(2\pi {\Theta_i \over {\Phi_0/2}}\right) +
{q_i^2\over 2C_{\rm t}} - {2e n_iq_i\over C_{\rm J}} + 
H_{\rm bath}(q_i) 
\right\} \nonumber \\
&& + {(\phi_1+\phi_2)^2\over 4L_a} + {(\phi_1-\phi_2)^2\over 4(2L+L_a)} \ . 
\end{eqnarray} 
Here $q_1$ is the charge on the left-most external capacitor, while $q_2$ 
is the charge on the right-most external capacitor. $\phi_1$
and $\phi_2$ are their conjugate phases (weakly fluctuating ones).
We make a canonical transformation
\begin{equation}
\label{2bit_Canonical_Dissipative}
 \begin{array}{ll}
   \tilde q_i = q_i - {2e n_iC_{\rm t}\over C_{\rm J}} \ , &  
   \tilde \phi_i = \phi_i \ ; \\
   \tilde n_i = n_i \ , &  
   \tilde \Theta_i = \Theta_i + {C_{\rm t}\over C_{\rm J}}\phi_i \ , \\
 \end{array} 
\end{equation}  
($i=1,\,2$) and the new Hamiltonian looks like (we omit the tildes):
\begin{eqnarray}
\label{2bit_Dissipative_Hamiltonian_Cosine}
H = && \sum_{i=1,2} \left\{ 
{(2e n_i - {CV_i/2})^2\over{C+2C_{\rm J}}} - 
E_{\rm J} \cos\left[{2\pi\over{\Phi_0/2}} \left(\Theta_i - 
{C_{\rm t}\over C_{\rm J}} \phi_i\right)\right] +
{q_i^2\over 2C_{\rm t}} + 
H_{\rm bath}\left(q_i +{C_{\rm t}\over C_{\rm J}}2e n_i\right) 
\right\} \nonumber \\ 
&& + {(\phi_1+\phi_2)^2\over 4L_a} + {(\phi_1-\phi_2)^2\over 4(2L+L_a)} \ .
\end{eqnarray}
Although the canonical transformation (\ref{2bit_Canonical_Dissipative})
looks equivalent to (\ref{Canonical_Transformation_Theta}),
there is one essential difference. In (\ref{Canonical_Transformation_Theta})
we used the phase variable which included the bath fluctuations,
while in (\ref{2bit_Canonical_Dissipative}) we keep the weakly
fluctuating phases $\phi_1$ and $\phi_2$. Next, we expand 
the $E_{\rm J} \cos(..)$ terms in $\phi_1$ and $\phi_2$ and we integrate over 
the baths and over the two oscillatory modes. Finally we assume 
$L_a \rightarrow 0$. One can directly check that in the nondiagonal 
channel the mixed term of (\ref{2bit_Coupling}) emerges, accompanied
with a bunch of other terms which are small at low frequency.
Thus, the two-bit coupling, crucial for the quantum computation,
is stable under the influence of the dissipation.

\section{Discussion}
Several conditions were assumed in this Letter in order to
guarantee the validity of our results. Here we state
these conditions once more and discuss the appropriate ranges
of parameters. We start with $E_{\rm J} \approx 1\,$K as a
suitable experimental condition. 
To satisfy $E_{\rm J} \ll E_{C_{\rm J}}$
we take $C_{\rm J} \approx 10^{-16}\, $F, which is an experimentally
accessible value. 
As we would like $A$ to be large (\ref{High_Frequency_Condition}), it 
seems that $L$ and $C_t$ should be as small as possible. 
However, the two-bit coupling energy, $E_L$ (\ref{E_L}),  
should be larger than the temperature of the experiment.
Assuming a reasonable working temperature of $20\,$mK, we demand 
$E_L \approx 0.1\,$K. {}From (\ref{High_Frequency_Condition}) and
(\ref{E_L}) we get $C_{\rm t}=E_L C_{\rm J}^2A^2/e^2$. To have a wide 
enough operation voltage interval we take $A\approx 10$, and 
obtain $C_{\rm t}\approx 10^{-17}-10^{-16}\,$F and 
$L \approx 10^{-8}-10^{-7}$H. Thus the renormalization
of $g$ is of the order of $10$, and $\tau_r/\tau_{\rm op}\approx 10^2-10^3$
(assuming the realistic value $R \approx 100\,\Omega$) 
(\ref{Energy_Relaxation_Time}). Finally we observe that in this range of 
parameters the inequalities (\ref{Small_Fluctuation_Assumption}) and 
(\ref{Frequences_Unequality}) are always satisfied. 
We conclude that the quantum manipulations we have discussed in this paper
can be tested experimentally using the currently available lithographic and
cryogenic techniques. Application of the Josephson junction system as an
element of a quantum computer is a more subtle issue, demanding either the 
fabrication of junctions with $C_{\rm J}<10^{-16}\,$F, or a further 
reduction of the working temperature.

\section{Acknowledgments:}
We thank T.~Beth, J.~E.~Mooij, A.~Zaikin and P.~Zoller for stimulating
discussions. This work is supported by the Graduiertenkolleg ''Kollektive
Ph\"anomene im Festk\"orper'',  by
the Sonderforchungsbereich 195 of the DFG and by 
the German Israeli Foundation (Contract G-464-247.07/95).

\end{document}